\documentclass[twocolumn,aps,prd,nofootinbib]{revtex4-1}
\usepackage[colorlinks=true,linkcolor=black,citecolor=blue,urlcolor=black]{hyperref}
\usepackage{graphicx,soul}
\usepackage{subfigure}
\usepackage{natbib}
\usepackage{epsfig}
\usepackage{amstext,amsmath,amsfonts,amssymb,amsthm}
\usepackage[format=plain,justification=centerlast]{caption}
\usepackage{color}

\usepackage{verbatim}
\usepackage{lipsum}
\usepackage{color,array,wrapfig}
\usepackage{caption}
\def\be{\begin{equation}}
\def\ee{\end{equation}}
\def\bes{\begin{eqnarray}}
\def\ees{\end{eqnarray}}
\def\ba{\begin{align}}
\def\ea{\end{align}}
\def\bwt{\begin{widetext}}
\def\ewt{\end{widetext}}

\def\aa{\alpha}

\newbox\one
\newbox\two
\long\def\loremlines#1{%
    \setbox\one=\vbox {%
       Test.\footnote{a footnote}%
      \lipsum\footnote{Another footnote.}%
     }
   \setbox\two=\vsplit\one to #1\baselineskip
   \unvbox\two}

\begin{document}
\title{The generation of helical magnetic field in a viable scenario of Inflationary Magnetogenesis}

\author{Ramkishor Sharma$^{1}$}
\email{rsharma@physics.du.ac.in, sharmaram.du@gmail.com}
\author{Kandaswamy Subramanian$^{2}$} 
\email{kandu@iucaa.in}
\author{T. R. Seshadri$^{1}$}
\email{trs@physics.du.ac.in} 
\affiliation{$^{1}$Department of Physics \& Astrophysics, University of Delhi, New Delhi$-$110007 India.}
\affiliation{$^{2}$ IUCAA, Post Bag 4, Pune University Campus, Ganeshkhind, Pune$-$411007 India.}

%
\begin{abstract}
\noindent  We study the generation of helical magnetic fields in a model of inflationary magnetogenesis which is free from the strong coupling and back-reaction problems. To generate helical magnetic fields, 
we add an $f^2 \tilde{F}^{\mu\nu} F_{\mu\nu}$ term to the lagrangian of Ratra model. The strong coupling and back-reaction problems are avoided if we take a particular behaviour of coupling function $f$, in which $f$ increases during inflation and decreases post inflation to reheating. The generated magnetic field is fully helical and has a blue spectrum, $d\rho_B/d\ln k \propto k^4$. 
This spectrum is obtained when coupling function $f\propto a^2$ during inflation. The scale of reheating in our model has to be lower than $4000$ GeV to avoid back-reaction post inflation. The generated magnetic field spectrum satisfies the $\gamma$-ray bound for all the possible scales of reheating. The comoving magnetic field strength and its correlation length are $\sim 4 \times 10^{-11} $ G and $70$ kpc respectively, if reheating takes place at 100 GeV. For reheating at the QCD scales of $150$ MeV, the field strength
increases to $\sim$ nano gauss, with coherence scale of $0.6$ Mpc.
\end{abstract}

\maketitle

\section{Introduction}
Cosmic magnetic fields have been detected from planetery scales \cite{planet} to galaxy clusters scales \cite{r-beck,clarke,widrow}. 
$\gamma$-ray observations of Blazars suggest their presence in the voids as well \cite{neronov,taylor2011}. However the origin of these fields do not have a fully satisfactory explanation. Astrophysical scenarios \cite{biermann,kandu1994,kulsrudbattery, zweibel2000, rees2006} for generating these fields involve
battery effects
to create a seed field
 which is later amplified to the observed strength by the dynamo action  \cite{zel, Shukurov,bks,kulsurd2008}. However, the presence of coherent magnetic fields in void regions favor primordial scanarios of generation. Several possible scenarios for this have been suggested in literature \cite{turner-widrow,vachaspati, ratra, Sigl:1996dm, kisslinger,dolgov2004, takahashi2005, shiv2005,qcd, martin-yokoyama, campanelli2008, rajeev2010, agullo2013,rajeev2013,caprini2014,kobayashi2014, atmjeet2014, atmjeet2015,sriram2015, Campanelli2015, 1475-7516-2015-03-040,sriram2016} (for reviews, see  \cite{grasso, dner, kandu2010, 2011PhR...505....1K, kandu2016}). 

Inflationary magnetogenesis is one of the possible scenarios to generate fields that are coherent over large scales. However the generation of magnetic field during inflation in the standard physics is not possible due to the conformal invariance of the electromagnetic field \cite{parker}. 
A breaking of conformal invariance is necessary. This has been done in many models by taking a time dependent function coupled with the kinetic energy term of the electromagnetic (EM) field. This time dependent coupling term can arise through the coupling between inflaton field and EM field (Ratra model) \cite{ratra} or by taking a non-minimal coupling of the EM field to the gravitational field \cite{turner-widrow}. 

Although Ratra model generates sufficient strength of the magnetic field, it potentially suffers from strong coupling and back-reaction problems \cite{mukhanov2009}. These problems have been resolved by further modification of the Ratra model \cite{rajeev2013,kobayashi2014, sharma2017}. In Ref. \cite{sharma2017}, to resolve these problems $f$ is assumed to increase during inflation and decrease back to its initial value post inflation. Firstly this behaviour of $f$ circumvents the strong coupling problem. Moreover, for a small enough inflationary and reheating scales, the model also does not suffer from the back-reaction problem. Indeed demanding no back-reaction post inflation, bounds on the inflationary scale and reheating scale has been obtained. The generated magnetic field strength in this model can also explain the magnetic field strength suggested by $\gamma$-ray observation, below a certain reheating scale. 

In the model discussed above, the generated magnetic field is of non-helical nature. In this paper we look at the possible
generation of helical magnetic fields. 
The non-linear evolution of magnetic field in helical case differs from non-helical one. Due to the helicity conservation, the magnetic field strength decreases at a slower rate and the correlation length increases at a higher rate compared to the non-helical case. Helicity conservation gives us a more optimistic
evolution of the magnetic field and this can relax the bound on the reheating scale given in \cite{sharma2017} further. 

Moreover, it has been claimed in the literature that gamma ray observations of the Blazars indicate the presence of helical magnetic field in intergalactic medium \cite{tashiro2014,tashiro2015}. To generate helical magnetic field, we add a parity breaking term in the Electromagnetic action. We also compare our generated magnetic field strength with the constraints from $\gamma$-ray observation.

This paper is organised as follows. In section {\ref{emdi} we discuss the generation of helical magnetic field during inflation. Section \ref{eainf} discusses the generation of the field post inflation to reheating and we obtain a relation between inflationary scale and reheating scale by demanding no back-reaction post inflation. Section \ref{eagen} incorporates the nonlinear
evolution of magnetic field and its correlation length after generation. 
Results of our model are 
also given there.
In section \ref{gamma}, we compare our results with the $\gamma$-ray observation. Our conclusions are presented in \ref{conclusion}.

\section{Generation of helical magnetic field during inflation}\label{emdi}
We start with the action for the electromagnetic field in which the conformal invariance is explicitly broken by introducing a time dependent function $f^2$ multiplying $F^{\mu \nu}F_{\mu \nu}$ in the action. To generate helical magnetic field, we also add a $f^2 {F_{\mu\nu}}{\tilde F^{\mu\nu}}$ in the action. Thus we take the action to be of the form,
\begin{eqnarray}
S &=-\int\sqrt{-g} d^4x [\frac{f^2(\phi)}{16 \pi}\left({F_{\mu\nu}}{F^{\mu\nu}}+{F_{\mu\nu}}{\tilde F^{\mu\nu}}\right) + j^{\mu}A_{\mu}]\nonumber\\
&-\int\sqrt{-g} d^4x \Big[\frac{1}{2}\partial^\nu \phi \partial_\nu \phi + V(\phi)\Big] \label{actionfull}
\end{eqnarray}
Here $F_{\mu\nu}=\partial_\mu A_\nu - \partial_\nu A_\mu$ and $\tilde F^{\mu\nu}=(1/2) \epsilon^{\mu\nu\aa\beta}F_{\aa\beta}$, where $A_\mu$ is the EM 4-potential and $\epsilon^{\mu\nu\aa\beta}$ is fully antisymmetric tensor defined as $\epsilon^{\mu\nu\aa\beta}=1/\sqrt{-g}~ \eta^{\mu\nu\aa\beta}$. $\eta^{\mu\nu\aa\beta}$ is levi-civita symbol whose values are $\pm1$ only. The term $j^\mu A_\mu$ represents the interaction where $j^\mu$ is the four current density. Remaining part of the action incorporates the evolution of the inflaton field. In this paper we adopt the following nomenclature. Greek indices $\mu, \nu....$ are for space-time coordinates and Roman indices $i,j,k....$ are for spatial coordinates. Our metric convention is $g_{\mu\nu}=diag(-,+,+,+)$. For further analysis we assume negligible free charge density during inflation. Hence, we neglect the interaction term. To obtain the equation of motion of 4-potential $A_{\mu}$ we vary the action with respect to the $A_{\mu}$.
\begin{align}
[f^2 \left(F^{\mu\nu}+\tilde F^{\mu\nu}\right)]_{;\nu}&=0\nonumber\\
\frac{1}{\sqrt{-g}}\frac{\partial}{\partial x^\nu}\Bigg[\sqrt{-g}f^2(\phi)\Big(g^{\mu\alpha}g^{\nu\beta}  F_{\alpha\beta}\nonumber\\
+\frac{1}{\sqrt{-g}}\eta^{\mu\nu\alpha\beta}F_{\alpha\beta}\Big)\Bigg]&=0 \label{ME}
\end{align}
Varying the action with respect to scalar field we obtain the equation governing the evolution of the scalar field as
\begin{align}
\frac{1}{\sqrt{-g}} \frac{\partial}{\partial x^\nu}\Big[\sqrt{-g} g^{\mu\nu}\partial_\mu \phi \Big] -\frac{dV}{d\phi}&=\frac{f}{2}\frac{df}{d\phi} \Big(F_{\mu\nu} F^{\mu\nu}\nonumber\\
&+F_{\mu\nu} \tilde F^{\mu\nu}\Big)
\end{align}
Here EM field is assumed to be a test field. Hence, it will not affect the geometry of spacetime. In the inflationary era the universe is dominated by scalar field $\phi$ which is a time dependent homogeneous field. We work in a frame work of homogeneous and isotropic universe, described by the  FRW line element.
\begin{eqnarray}
ds^2&=&-dt^2+a^2(t)[dx^2+dy^2+dz^2]\nonumber\\
&=& a^2 (\eta)[-d\eta^2+dx^2+dy^2+dz^2]
\end{eqnarray}
In this new coordinate system $(\eta,x,y,z)$, we can define fundamental observers with four velocity $(\frac{1}{a},0,0,0)$. It is convenient to work in Coulomb gauge,
$$\partial_j A^j =0 \quad , \quad A_0=0$$
For $\mu=i$, Eq. (\ref{ME}) is
\begin{equation}\label{A}
A_i''+2 \frac{f'}{f} \left(A_i'+\eta_{ijk}\partial_j A_k\right)-a^2 \partial_j \partial^j A_i=0
\end{equation}
Here prime(') denotes derivative with respect to $\eta$ and $\partial^j$ is defined as $\partial^j\equiv g^{jk}\partial_k=a^{-2 }\delta^{jk} \partial_k$. $\eta_{ijk}$ represents the three dimensional levi-civita symbol. To quantize the EM field, we calculate the conjugate momentum ($\Pi^i$)of $A^i$ field, promote these fields to operators and impose the canonical quantization condition,
\begin{align}
\Pi^i=\frac{\delta L}{\delta \dot A_i},
\end{align}
\begin{align}
[A^i,\Pi_j]=i \int \frac{d^3k}{2\pi^3} e^{\vec{k} \cdot (\vec{x}-\vec{y})} P^i_j (k).
\end{align}
Here $P^i_j=\delta^i_j-\delta_{jm}(k^ik^m/k^2)$ is used to ensure the coulomb gauge condition.
We fourier transform $A_{i}$ using the momentum space operators ($b_\lambda$ and $b_\lambda^{\dagger}$).
\begin{align}
A^i(\vec{x},\eta)&= \sqrt{4\pi}\int \frac{d^3 k}{2 \pi^3} \sum_{\lambda=1,2}\epsilon^{i}_{\lambda}\Big[A_\lambda(k,\eta) b_\lambda(\vec{k}) e^{i\vec{k}\cdot \vec{x}}\nonumber\\
&+A_\lambda^*(k,\eta) b^{\dagger}_\lambda(\vec{k}) e^{-i\vec{k}\cdot \vec{x}}\Big]. \label{fta}
\end{align}
Here $\epsilon^{i}_{\lambda}$ represents the polarisation vector which can be written in terms of the 3-dimensional orthonormal unit vectors as
\begin{align}
\epsilon^{i}_{\lambda}=\frac{\hat{\epsilon}^{i}_{\lambda}}{a}~~,~~\epsilon^{i}_{3}=\frac{\hat{k}}{a}
\end{align}
Here $\hat{\epsilon}^{i}_{\lambda}$ are unit 3-vectors, orthogonal to $\hat{k}$ and each other.
It is useful to define a new variable $\bar {A}_\lambda$ as, $\bar A_\lambda =a A_\lambda(k,\eta)$. Substituting Eq.(\ref{fta}) in Eq.(\ref{A}).
\begin{equation}
\sum_\lambda b_{\lambda}\Big[\hat{\epsilon}_{i\lambda}\left(\bar{A}_\lambda''+2 \frac{f'}{f} \bar{A}_\lambda'+k^2 \bar{A}_\lambda\right)+\frac{2 f'}{f}\eta_{ijm}\hat{\epsilon}_{m\lambda}k_j\bar{A}_\lambda\Big]=0\label{AA}
\end{equation}
To simplify this equation further let's choose a different set of basis vector defined as $\hat{\epsilon}_+=(\hat{\epsilon}_1+i\hat{\epsilon}_2)/2$ and $\hat{\epsilon}_-=(\hat{\epsilon}_1-i\hat{\epsilon}_2)/2$. In terms of these new basis
vectors, 
$\sum_{\lambda} \bar{A}_\lambda \hat{\epsilon}_\lambda b_\lambda=\bar{A}_+ \hat{\epsilon}_+b_++\bar{A}_-\hat{\epsilon}_- b_-$. This set of basis 
vectors are known as the helicity basis.
Then Eq.(\ref{AA})
reduces to 
\begin{equation}
\bar{A}_h''+2 \frac{f'}{f} \left(\bar{A}_h'+h k \bar{A}_h\right)+k^2 \bar{A}_h=0 \label{genA}
\end{equation}
Here $h=\pm 1$ represents the helicity sign. 
The equation of motion in terms of $\mathcal A_h =f \bar{A}_h(k,\eta)$ turns out to be,
\begin{equation}
\mathcal A_h''(k,\eta)+\Big(k^2-\frac{f''}{f}+2 h k \frac{f'}{f} \Big) \mathcal A_h(k,\eta)=0   \label{scriptA}
\end{equation}
Before we solve the above equation for a particular $f(\phi)$, it will be of interest to calculate the energy density of the EM field. To do this we calculate the energy momentum tensor of the EM field.
\begin{align}
T_{\mu\nu}&\equiv-\frac{2}{\sqrt{-g}}\frac{\delta \sqrt{-g} \mathcal{L}}{\delta g^{\mu\nu}}\nonumber\\
&=\frac{f^2}{4\pi}\left[g^{\alpha\beta}F_{\mu\alpha}F_{\nu\beta}-g_{\mu\nu}\frac{F^{\alpha\beta}F_{\alpha\beta}}{4}\right]
\end{align}

We define the electric and magnetic field four vectors as
$B^\mu\equiv \tilde{F}^{\mu\nu}u_\nu$ and $E^\mu\equiv F^{\mu\nu}u_\nu$. 
For the observer $u^\mu=(1/a,0,0,0)$, the time component of these vectors are zero. The spatial components are given by $B_i=(1/a) \eta_{ijk}\delta^{jm}\delta^{kn}\partial_m A_n$ and $E_i=-(1/a) \partial_\eta A_i$.
Then the
EM energy densities of ground state measured by the observer with 4-velocity $u^\mu=(1/a,0,0,0)$ are given by,
\begin{align}
\rho_B = \langle0|T^{B}_{\mu \nu}u^{\mu}u^{\nu}|0\rangle \quad \text{and} \quad \rho_E = \langle0|T^{E}_{\mu \nu}u^{\mu}u^{\nu}|0\rangle ,\label{rhobae}
\end{align}
where 
we have separated the total energy density into the magnetic part and the electric part. We express these parts in terms of $A^i$ as,
\begin{eqnarray}
\rho_B &=&\langle0|\frac{f^2}{8 \pi} \left[\partial_i A_n \partial_j A_l \left(g^{ij}g^{nl}-g^{il}g^{nj}\right)\right]|0\rangle\\ \label{rhob}
&=&\langle0|\frac{f^2}{8 \pi} B^i B_i|0\rangle\nonumber\\
\text{and} \quad \rho_E &=& \langle0| \frac{f^2}{8 \pi} \left[A'_i A'_j  g^{ij}\right]|0\rangle\\
&=&\langle0|\frac{f^2}{8 \pi} E^i E_i|0\rangle\nonumber \label{rhoe}
\end{eqnarray}

After substituting $A_i$ from Eq.(\ref{fta}) into Eq.(\ref{rhob}) and Eq.(\ref{rhoe}), and using the helicity basis, we reduce the energy densities in terms of $\mathcal{A_+}$ and $\mathcal{A_-}$.
\begin{eqnarray}
\rho_B\equiv\int \frac{dk}{k}\frac{d\rho_B(k,\eta)}{d \ln k} &=&\int \frac{dk}{k}\frac{1}{(2 \pi)^2} \frac{k^5}{a^4} \Big(|\mathcal A_+(k,\eta)|^2\nonumber\\
&+&|\mathcal A_-(k,\eta)|^2\Big) \label{srhob}\\
\rho_E\equiv\int \frac{dk}{k}\frac{d\rho_E(k,\eta)}{d \ln k} &=&\int \frac{dk}{k}\frac{f^2}{(2 \pi)^2} \frac{k^3}{a^4} \Bigg(\left|\Big[\frac{\mathcal A_+(k,\eta)}{f}\Big]'\right|^2\nonumber\\
&+&\left|\Big[\frac{\mathcal A_-(k,\eta)}{f}\Big]'\right|^2\Bigg) \label{srhoe}
\end{eqnarray}
In deriving the above expressions we have used the following properties,
\begin{align}
b_h|0\rangle=0~~~~\langle0|b_h (\vec{k}) b^{\dagger}_{h'}(\vec{k'})|0\rangle=(2\pi)^3 \delta_{h h'}\delta^3(\vec{k}-\vec{k'})\nonumber
\end{align}
In our analysis, we work with the spectral energy densities of magnetic and electric fields given by
$(d\rho_B(k,\eta)/d\ln k)$ and $(d\rho_E(k,\eta)/d\ln k)$ respectively. These spectral densities represent the energy contained in logarithmic interval in k-space.

We now turn to calculation of the magnetic and electric spectral energy densities for a particular type of coupling function. Let us assume the form of the coupling function to be a simple power law as,
\begin{equation}
f_1(a) = f_i\left(\frac{a}{a_i}\right)^{\aa}.\label{ff}
\end{equation}
We assume that during inflation $f(\phi)$ has a form such that $f$ evolves with $a$ as given in $\eqref{ff}$. Here $a_i$ represents the value of scale factor at the beginning of inflation and $\aa$ is a real constant. By assuming the background to be purely de-sitter during inflation ($a\propto \eta^{-1}$), we get the following
\begin{equation}
\frac{f_1''}{f_1}=\frac{\aa(\aa+1)}{\eta^2}.
\end{equation}
For this form of coupling function Eq.(\ref{scriptA}) reduces to
\begin{equation}
\mathcal A_h''(k,\eta)+\Big(k^2-\frac{\aa (\aa+1)}{\eta^2}-2 h k \frac{\aa}{\eta} \Big) \mathcal A_h(k,\eta)=0.   \label{scriptAf}
\end{equation}
To solve Eq.(\ref{scriptAf}), we rewrite it by defining some new variables, $\mu^2\equiv \aa(\aa+1)+\frac{1}{4}, \quad \kappa \equiv i \aa h, \quad z\equiv 2 i k \eta $. Then the equation takes the form,
\begin{equation}
\frac{\partial ^2 \mathcal A_h(k,\eta)}{\partial z^2}+\left[\frac{1}{z^2}\left(\frac{1}{4}-\mu^2\right)+\frac{\kappa}{z}- \frac{1}{4} \right] \mathcal A_h(k,\eta)=0.   \label{scriptAfnew}
\end{equation}
The solutions of this equation are Whittaker functions \cite{whittakar},
\begin{equation}
\mathcal A_h = c_1 W_{\kappa,\mu}(z)+c_2 W_{-\kappa,\mu}(-z). 
\end{equation}
To determine the coefficients $c_1$ and $c_2$, we have matched the solution with the bunch davies vacuum in the sub horizon limit $|-k\eta|>>1$, $\mathcal{A}_h=\frac{e^{-i k \eta}}{\sqrt{2 k}}$. After matching we get,
$$\mathcal A_h = \frac{e^{\frac{i \pi \kappa}{2}}}{\sqrt{2 k}} W_{\kappa,\mu}(z)= \frac{e^{\frac{-h \pi \aa }{2}}}{\sqrt{2 k}} W_{i \aa h, \aa +\frac{1}{2}}(2 i k \eta)$$
At the end of the inflation, all the modes of cosmological interest will be outside the horizon. To get the spectral magnetic energy density of these modes, we need to write the above expression in the super horizon limit.
\begin{align}
\mathcal A_h&=\frac{C_h}{\sqrt{2k}} \Big[(-k \eta)^{-\aa}+h(-k \eta)^{-\aa+1}\Big]+\nonumber\\
&~~~\frac{D_h}{\sqrt{2k}} \Big[(-k \eta)^{1+\aa}-\frac{h \aa}{1+\aa}(-k\eta)^{2+\aa}\Big] \label{sa}
\end{align}
Here 
\begin{align}
C_h&=e^{\frac{-h\pi \aa}{2}}\frac{(-2 i)^{-\aa } \Gamma (1+2 \aa) }{\Gamma (1+\aa + i h \aa)},\nonumber\\
D_h&=e^{\frac{-h\pi \aa}{2}}\frac{(-2 i)^{1+\aa } \Gamma (-1-2 \aa) }{\Gamma (-\aa - i h \aa)}.
\end{align}

Substituting Eq.(\ref{sa}) into Eq.(\ref{srhob}) the spectral magnetic energy density comes out to be,
\begin{eqnarray}
\frac{d\rho_B}{d \ln k}&=&\frac{1}{8 \pi^2} H_f^4 \Big[\left(|C_+|^2+|C_-|^2\right)(-k \eta)^{-2\aa+4}\nonumber\\
&+&\left(|D_+|^2+|D_-|^2\right)(-k \eta)^{2\aa+6}\Big].  \label{rhob2}
\end{eqnarray}
Similarly the spectral electric energy density is given by,
\begin{eqnarray}
\frac{d\rho_E}{d \ln k}&=&\frac{1}{8 \pi^2} H_f^4 \Big[\left(|C_+|^2+|C_-|^2\right)(-k \eta)^{-2\aa+4}\nonumber\\
&+&\left(|D_+|^2+|D_-|^2\right)(1+2\aa)^2(-k \eta)^{2\aa+4}\Big].  \label{rhoe2}
\end{eqnarray}
For the expressions in Eq.(\ref{rhob2}) and Eq.(\ref{rhoe2}), we have only kept the dominant terms. 

The reason of two branches in the above expressions is that $\aa$ can be positive or negative. In the magnetic field spectrum the first branch dominates when $\aa>1/2$ and the other branch dominates when $\aa<1/2$. For the electric field spectrum the first branch dominates when $\aa>-1/2$ and the other branch dominates when $\aa<-1/2$.
There are two possible value of $\aa$ (namely, $\aa=2,-3$) for which the magnetic field spectrum is scale invariant.

For the case $\aa=-3$, when the magnetic field is scale invariant the electric field spectrum diverges as $(-k\eta)^{-2}$. This implies that the electric field energy density may overshoot the inflaton energy density during inflation in this case and our assumption of EM field being a test field would no longer be valid. This problem is known as the back-reaction problem \cite{mukhanov2009}.

However, for the case $\aa=2$, both the magnetic and electric field spectrum are scale invariant, and this avoids the back reaction problem. 
For this case the coupling function $f$ is proportional to $a^2$, which means that $f$ will be very large at the end of inflation compared to its initial value. If we assume that $f$ becomes a constant at the end of inflation, then the effective EM charge $e_f=1/f^2$.
Suppose we demand that at the end of inflation $e_f$ should have the observed value, then it will be very large at the beginning of inflation. Due to this large value of $e_f$, our perturbative analysis of field theory would no longer be valid. This problem is known as strong coupling problem \cite{mukhanov2009}. 
On the other hand, if $e_f$ has the observed value at the beginning of
inflation, it 
 will have a very small value at the end of inflation, avoiding the strong coupling problem. This case is further explored 
in the next section.



The branch $\aa=2$ is also preferred, as we discussed in detail in Ref. \cite{sharma2017}, because it evades the constraints imposed by the possibility of increased conductivity due to the Schwinger effect \cite{kobayashi2014}. 
The magnetic energy spectrum at the end of inflation for $\aa=2$,
\begin{align}
\frac{d\rho_B}{d \ln k}&\approx \frac{9~ e^{4\pi}}{320 \pi^3} H_f^4
\end{align}
We note that this value is larger than the non-helical case by a factor of $(e^{4 \pi}/80 \pi) \approx 10^3$.
\section{Evolution after inflation}\label{eainf}
As discussed in the last section, there are two possible scenarios for obtaining a scale invariant magnetic spectrum. The case in which $f$ is increasing during inflation avoids the back-reaction problem. Moreover if we assume that $f$ begins with a value of unity at the onset of inflation and increases during the inflationary phase, there will be no strong coupling problem. At end of inflation, however, the EM field will be very weakly coupled to the charged particles as $f$ is much larger than unity. We address this issue in \cite{sharma2017} by postulating that from the end of inflation onwards $f$ decreases and attains a value of unity at reheating and remains unity thereafter. This ensures that EM action is again in the standard conformally invariant form after the reheating era. Thus the deviation from the standard form is onset of inflation to the reheating.

We assume the universe to be matter dominated from the end of inflation to reheating. We consider $\aa > 1/2$ for further analysis. In this era the evolution of the scale factor is as follows, $$a=\frac{a_f^3 H_f^2}{4}\left(\eta+\frac{3}{a_f H_f}\right)^2$$
and the coupling function $f$ is assumed to evolve as
\begin{align}
f\propto\left(\frac{a}{a_f}\right)^{-\beta}\nonumber
\end{align}
Here $a_f$ is the value of scale factor at the end of inflation. We calculate the constant of proportionality by demanding the continuity of $f$ at the end of inflation.
\begin{align}
f_2=f_i\left(\frac{a_f}{a_i}\right)^{\aa}\left(\frac{a}{a_f}\right)^{-\beta}\nonumber
\end{align}
To estimate the EM energy densities in this era, we need to solve Eq.(\ref{scriptAf}) for this new coupling function. Solution is given by,
\begin{equation}
\mathcal A_{2h} = d_1 M_{2 i \beta h,-(2 \beta +\frac{1}{2})}(2 i k \zeta)+ d_2 M_{2 i \beta h, 2 \beta +\frac{1}{2}}(2 i k \zeta)\label{ainfa}
\end{equation}
Here $\zeta=\eta+3/(a_f H_f)$ and $M_{2 i \beta h,-(2 \beta +\frac{1}{2})}(2 i k \zeta)$ represents the second kind of whittaker function \cite{whittakar}. To calculate $d_1$ and $d_2$, we need the above expression in the super horizon limit.
In this limit Eq.($\ref{ainfa}$) becomes,
\begin{align}
\mathcal A_{2h} &=d_1 (2 i)^{-2\beta}\left[(k \zeta)^{-2\beta}- h(k \zeta)^{-2\beta+1}\right]\nonumber\\
&+d_2 (2 i)^{2\beta+1}\left[(k\zeta)^{2\beta+1}+\frac{2 h \beta}{1+2\beta}(k\zeta)^{2\beta+2}\right]\nonumber
\end{align}
and, 
\begin{align}
\bar{A}_{2h}=\frac{\mathcal A_{2h}}{f_2}&=\left(\frac{k}{H_f}\right)^{-2\beta}\Bigg(d_3 \left(1- h(k \zeta)\right)\nonumber\\
&+d_4 \left((k\zeta)^{4\beta+1}+\frac{2 h \beta}{1+2\beta}(k\zeta)^{4\beta+2}\right)\Bigg)
\end{align}
Here $d_3$ and $d_4$ are two new constants. They can be expressed in terms of $d_1$ and $d_2$. We demand that at the end of the inflation both $\bar{A}_h$, $\bar{A}_{2h}$ and their derivatives have to be matched.
After matching we get,
\begin{align}
d_3&= \frac{C_h}{\sqrt{2 k}} \left(\frac{k}{H_f}\right)^{-\aa+2\beta}\left(1+3h\left(\frac{k}{a_f H_f}\right)\right)\nonumber
\end{align}
 and
\begin{align}
d_4&=\frac{C_h}{\sqrt{2 k}} \left(\frac{k}{H_f}\right)^{-\aa+2\beta} \frac{3 h^2}{2(4\beta+1)} \left(\frac{2 k}{a_f H_f}\right)^{-4\beta+1} \nonumber\\
&\left(1-\frac{2 h \beta}{1+2\beta} \frac{4\beta+2}{4\beta+1} \left(\frac{2 k}{a_f H_f}\right)\right)^{-1}\nonumber
\end{align}
In $d_3$ and $d_4$ expressions, we only take the contribution of dominant terms. 
Energy densities after inflation evolve as
\begin{align}
\frac{d\rho_B}{d \ln k}&=\frac{C_1}{8 \pi^2} \frac{k^4}{a^4} f_2^2(a) \left(\frac{k}{H_f}\right)^{-2\aa} \Bigg(1+ \frac{9}{4(4\beta+1)^2}\nonumber\\
& \left(\frac{2 k}{a_f H_f}\right)^{-8\beta+2} \left(\frac{2k} {a H}\right)^{8\beta+2}+\frac{3 h^2}{(4\beta+1)} \nonumber\\
&\left(\frac{2 k}{a_f H_f}\right)^{-4\beta+1} \left(\frac{2k} {a H}\right)^{4\beta+1}\Bigg)\nonumber
\end{align}
\begin{align}
\frac{d\rho_E}{d \ln k}&=\frac{1}{8 \pi^2} \frac{k^4}{a^4}f_2^2(a)\left(\frac{k}{H_f}\right)^{-2 \aa} \Bigg(C_1 + \frac{9}{4} C_1 \nonumber\\
& \left(\frac{2 k}{a_f H_f}\right)^{-8\beta+2} \left(\frac{2k} {a H}\right)^{8\beta}+3 C_2  \nonumber\\
&\left(\frac{2 k}{a_f H_f}\right)^{-4\beta+1}\left(\frac{2k} {a H}\right)^{4\beta}\Bigg)\nonumber
\end{align}
Here $C_1=\left|C_+\right|^2 +\left|C_-\right|^2$ and $C_2=\left|C_+\right|^2 -\left|C_-\right|^2$. At the end of inflation, the first term inside the bracket in the expressions of $d\rho_B/d \ln k$ dominates for all the modes outside the horizon and gives a scale invariant magnetic field spectrum for $\aa=2$. As $f$ decreases post inflation, this term also decreases and becomes very small at reheating. Although the second and third term are very small compared to the first term for the mode $k_i=a_i H_f$ at the end of inflation, the contribution from these terms compared to the first term increases as $f$ decreases post inflation. Consequently, the second and third term overshoot the first term before reheating. The second term is $36/((4 \beta +1)^2 a_f^4)\times(a_r/a_f)^{4\beta+1}$ times larger than the first term  and $6/((4 \beta +1) a_f^2)\times(a_r/a_f)^{2\beta+1/2}$ times larger than the third term at reheating for the mode which exit the horizon at the beginning of inflation ($k_i=a_i H_f$). It is even larger for all other modes of interest. After taking the contribution of the dominant term at reheating in the above expressions, we get the following expression for EM energy densities at reheating,
\begin{align}
\frac{d\rho_B}{d \ln k}\Bigg|_r&=\frac{9}{32 \pi^2} \frac{k^4}{a_r^4} f_2^2(a_r) \left(\frac{k}{H_f}\right)^{-2\aa} C_1\nonumber\\
&\frac{1}{(4\beta+1)^2} \left(\frac{2 k}{a_f H_f}\right)^{-8\beta+2} \left(\frac{2k} {a_r H_r}\right)^{8\beta+2}
\end{align}
\begin{align}
\frac{d\rho_E}{d \ln k}\Bigg|_r&=\frac{9}{32 \pi^2} \frac{k^4}{a_r^4}f_2^2(a_r)\left(\frac{k}{H_f}\right)^{-2 \aa}C_1\nonumber\\
&\left(\frac{2 k}{a_f H_f}\right)^{-8\beta+2} \left(\frac{2k} {a_r H_r}\right)^{8\beta}
\end{align}
Here $a_r$ and $H_r$ are the scale factor and Hubble parameter at reheating respectively.

Fig.~(\ref{final plot}) shows the evolution of EM and the inflaton 
energy densities with the scale factor, both during and after inflation. 
From Fig.(\ref{final plot}) one can see that EM energy densities are increasing after inflation. It is necessary that EM energy density does not overshoot the energy density of the universe before the coupling function $f$ reaches to its pre-inflationary value. Since EM energy density has a monotonically increasing behaviour, if $\rho_E+\rho_B< \rho_{\phi}$ is satisfied at reheating, it will be valid throughout the post inflationary era prior to reheating. The total EM energy density at reheating is,
\begin{widetext}
\begin{align}
 \rho_E+\rho_B \Big|_r&=\int_{a_i H_f}^{k_r} d \ln k \left(\frac{d \rho_E(k,\eta)}{d \ln k}\Big|_r +
\frac{d\rho_B(k,\eta)}{d\ln k}\Big|_r\right)  \nonumber\\
&=\int_{a_i H_f}^{k_r} d \ln k \Bigg[\frac{9}{32 \pi^2} \frac{k^4}{a_r^4}f_2^2(a_r)C_1\left(\frac{k}{H_f}\right)^{-2 \aa} \left(\frac{2 k}{a_f H_f}\right)^{-8\beta+2} \left(\frac{2k} {a_r H_r}\right)^{8\beta}\left(1+\frac{1}{(4\beta+1)^2}\left(\frac{2k} {a_r H_r}\right)^2\right)\Bigg]\nonumber\\
&=\frac{9}{32 \pi^2} \frac{f_2^2(a_r)}{a_r^4}C_1\left(\frac{k_r}{H_f}\right)^{-2 \aa}\left(\frac{2 k_r}{a_f H_f}\right)^{-8\beta+2} \left(\frac{2k_r} {a_r H_r}\right)^{8\beta} k_r^4 \left(\frac{1}{6-2\aa}+\frac{1}{(8-2\aa)(4\beta+1)^2}\left(\frac{2k_r} {a_r H_r}\right)^2\right) \label{intrho}
\end{align}
\end{widetext}
For further analysis we have used two new variable defined as,
$$N=\ln\left(\frac{a_f}{a_i}\right)~~~~~~ \text{and}~~~~~ N_r=\ln\left(\frac{a_r}{a_f}\right)$$
After substituting $k_r=a_r H_r=a_f H_f e^{-N_r/2}$ in Eq.($\ref{intrho}$) and using the definition of $N$ and $N_r$, we get
\begin{align}
 \rho_E+\rho_B \Big|_r&=C_3 H_f^4 e^{\aa(2N+N_r)-7 N_r}
 \end{align}
 Here $$C_3=\frac{9C_1}{8 \pi^2}\left(\frac{1}{6-2\aa}+\frac{4}{(8-2\aa)(4\beta+1)^2}\right).$$ In the above expression, we also use $\beta=\aa N/N_r$ which is obtained by demanding $f(a_r)=1$. 

 \begin{figure}[!]
 \epsfig{figure=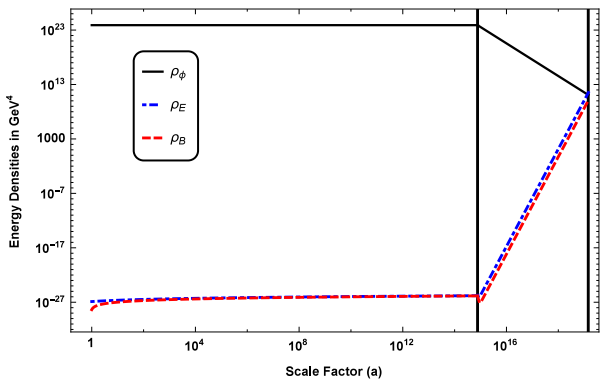,height=5cm,width=8cm,angle=0}
  \caption{In this figure we have taken $\aa=2$ and $T_r=100$ GeV. It shows the evolution of $\rho_{\phi}$ , $\rho_E$ and $\rho_B$ with scale factor. First vertical bold black lines is for the value of $a_f$ and second is for the value of $a_r$. This figure shows that the energy of EM field does not overshoot the energy of the scalar field $\phi$ which decides the background geometry if the scale of inflation and reheating satisfies the bound in Eq.($\ref{bound}$).}
  \label{final plot}
\end{figure}

In order that $\rho_E+\rho_B|_r< \rho_{\phi}|_r$, we require
\begin{align}
2\aa(N+N_r)-(7+\aa)N_r<\ln\left(\frac{\pi^2 g_r}{ 30 C_3}\right)-4\ln\frac{H_f}{T_r}.\label{bound1}
\end{align}
Here we use $\rho_\phi|_r= g_r (\pi^2/30)T_r^4$ where $T_r$ and $g_r$ represent reheating temperature and relativistic degree of freedom, respectively, at reheating. We have several variables in the above expression but they are all not independent. To reduce the expression in terms of the minimum number of variables (independent variables), we use the following constraint.

From the isotropy of the cosmic microwave background radiation, we find the following relation.
\begin{align}
 N+N_r&>66.9-\ln\left(\frac{T_r}{H_f}\right)-\frac{1}{3}\ln\frac{g_r}{g_0}.\label{nnr}
\end{align}
Here, $ g_{0} $ is the relativistic degree of freedom in the universe at present. The above relation has been derived from the fact that the present observable universe has to be inside the Hubble radius at the beginning of inflation. In the above expression, we also assume a radiation dominated era from reheating till today.

By the assumption of matter dominance post inflation to reheating, $N_r$ can be written in terms of $H_f$ and $T_r$.
\begin{align}
N_r=\frac{1}{3} \ln \frac{\rho_{\phi}|_{inf}}{\rho_{\phi}|_r}=\frac{1}{3} \ln\left[\frac{90 H^2_{f}}{8 \pi G \pi^2 g_r T_r^4}\right]\label{nr}
\end{align}
Substituting Eq.$\eqref{nnr}$ and Eq.$\eqref{nr}$ into Eq.\eqref{bound1} and writing $N_r$ in terms of $H_f$ and $T_r$, the bound in Eq.\eqref{bound1} reduces to 
\begin{align}
&\ln\Bigg[\frac{C_3}{ g_r}\left(\frac{g_0}{g_r}\right)^{\frac{2\aa}{3}}\left(\frac{g_r \pi^2}{30}\right)^{\frac{7+\aa}{3}}\Bigg]+134\aa+(2\aa+4)\ln\frac{H_f}{T_r}\nonumber\\
&-\frac{4(7+\aa)}{3}\ln\left(\sqrt[4]{\frac{3 H_f^2}{8\pi G}}\frac{1}{T_r}\right)<0.  \label{bound}
\end{align}
If reheating temperature and the scale of inflation satisfy the above bound, there will not be any back-reaction and strong coupling problem in our prescribed model till reheating.

For further analysis, we assume a particular value of $\aa$ and calculate the possible inflationary scales ($H_f$) for different reheating scales ($T_r$) using the bound in Eq. ($\ref{bound}$). For these $T_r$ and $H_f$, we calculate $N$ and $N_r$. We have also calculated correlation length of magnetic field and its strength at this scale at reheating using the following expressions,
\begin{align}
L_c&= a_r \frac{\int_{0}^{k_r} \frac{2 \pi}{k} \frac{d \rho_B(k,\eta)}{d \ln k} {d \ln k}}{\int_{0}^{k_r} \frac{d \rho_B(k,\eta)}{d \ln k} d \ln k}\nonumber\\
B[L_c]&= \sqrt{8 \pi \frac{d \rho_B(k,\eta)}{d \ln k}}\Big|_{k=\frac{2\pi a_r}{L_c}}.
\label{Lc}
\end{align}
 \begin{figure*}
\epsfig{figure=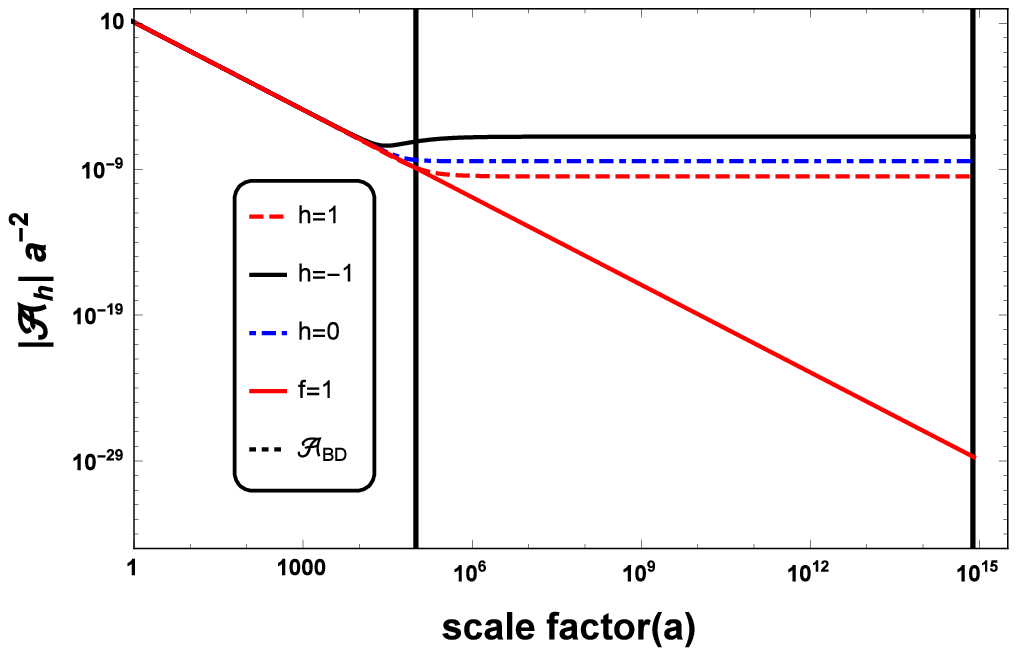,height=5cm,width=8cm,angle=0}
\epsfig{figure=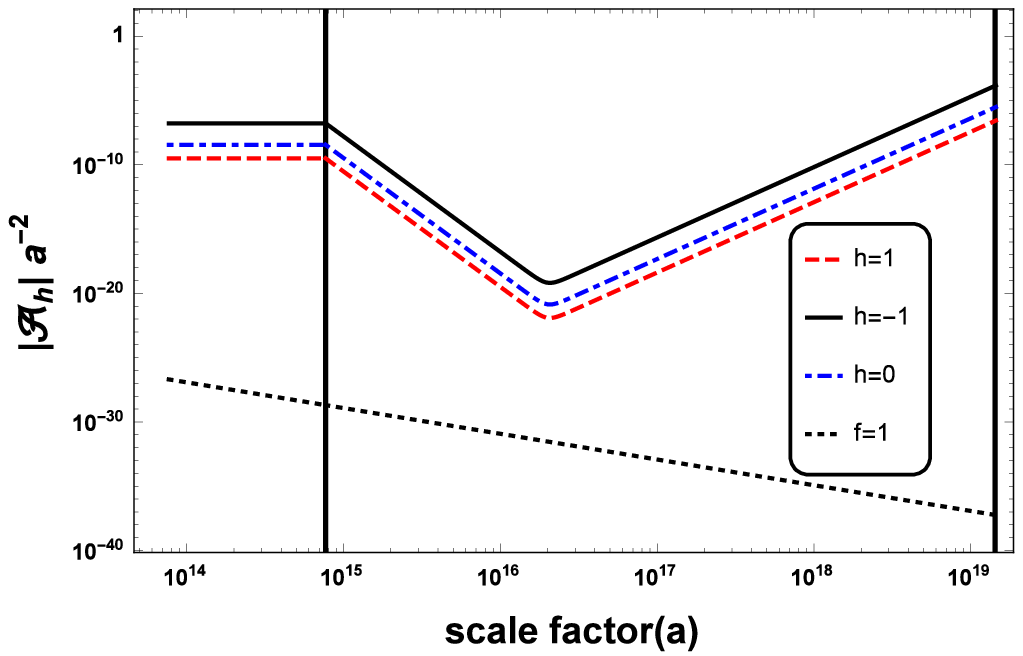,height=5cm,width=8cm,angle=0}
  \caption{In this figure we have plotted the $\mathcal{|A}_h|/a^2$ vs scale factor ($a$). Here we have assumed $\aa=2$, $T_r=100$GeV and $k=10^{5} H_f$.  In the left panel, we have shown that how vector potential evolves for positive helicity, negative helicity, zero helicity and  for the case of constant coupling ($f=1$) during inflation. In the second plot, we have shown the evolution of the same modes post inflation to reheating.}
\label{figvec}
\end{figure*}

In Fig.(\ref{figvec}), we plot the evolution of the $\mathcal{|A}_h|/a^2$,
(which appears in Eq.~(\ref{srhob}) for $\rho_B$),
with scale factor for different helcity modes from the beginning of inflation to the epoch of reheating. The  black solid curve and red dotted curve shows the evolution of the negative helicity and positive helicity mode, respectively. Blue dot-dashed curve shows the evolution of $\mathcal{|A}_h|/a^2$ if the parity breaking term is not present in the action (non-helical case) and for future purpose we name this mode as zero helicty mode. If $f$ had been a constant equal to 1, red solid curve would have represent the evolution. Black dotted curve shows the $\mathcal{|A}_h|/a^2$ for Bunch Davies vacuum. In the left panel, first vertical line is for the epoch of horizon crossing during inflation and second vertical line is for the end of inflation. It is evident from the figure that negative helicity mode has larger value than the zero helicty mode and the positive helicity mode. This means that in the helical case, magnetic energy density is larger compared to the non-helical case at the end of inflation and it is almost fully helical because the strength of positive helicity mode is negligible compared to negative helicity mode. It is also evident from this panel that without the coupling between inflaton and EM field, the strength of the magnetic field is very small for the modes which have crossed the horizon much earlier than end of the inflation.

The right panel of Fig.(\ref{figvec}) also follows the same colour coding and it shows the post inflationary evolution till reheating. In this panel, first vertical line is for epoch of end of inflation and the second vertical line is for the epoch of reheating. This panel shows that the strength of the mode decreases post inflation. Subsequently there is a transition and the mode starts to increase till reheating. The reason for the transition is as follows. The branch which dominated during inflation leads to both an initially dominant decaying mode and a subdominant growing mode after the transition to the matter dominated era post inflation. The initially dominant mode decreases as $f$ decreases, while the initially subdominant one increases with time. In further evolution, naturally there is a point where both the branch crosses each other, this point is the transition point in the right panel. Subsequent growth of the field continues till the reheating epoch (indicated by the second vertical line). Finally on reheating the electric field gets damped by the increased plasma conductivity. 
The magnetic field evolves further as discussed below.

\section{Evolution of magnetic field after reheating}\label{eagen}
To determine the magnetic field strength and its correlation length at present, we need to evolve the magnetic field from the epoch of reheating to today. As we have seen in the last section, our generated magnetic field has a blue $k^4$ spectra on super horizon scales at the time of generation. After reheating, the universe is dominated by radiation and in radiation dominance Hubble radius increases faster than the wavelength of a mode. Due to this modes start to re-enter the horizon. As the Alfv\'en crossing time for a mode becomes smaller than the comoving Hubble time, non-linear effects due to the magnetic field coupling to the plasma come into picture.

If we consider only the flux frozen evolution of the magnetic field $B\propto1/a^2$, then magnetic field strength and its correlation length at present are given by the following expressions.
\begin{align}
L_{c0}=& L_c\left(\frac{a_0}{a_r}\right), \nonumber \\ 
B_0[L_{c0}]=& B[L_c]\left(\frac{a_0}{a_r}\right)^{-2}.  
\label{BL_linear}
\end{align} 
However if we incorporate the non-linear effects and the consequent turbulent decay of the magnetic field, its strength and correlation scale have different scaling behaviour. Because of magnetic helicity conservation, inverse cascade takes place and magnetic energy transfers from smaller length scales to larger scales. This phenomenon has been discussed   in \cite{jedamzik, kandu2016, axel2001} and also confirmed by the numerical simulations \cite{axel2001,jedamzik,tina2013,axel2016,axel2017,axel2017-1}. After using the results discussed in \citep{kandu2016,axel2001,jedamzik}, 
we get the following scaling laws for the correlation scale $L_{c0}^{NL}$ and the strength of the field at this scale $B^{NL}_0[L_{c0}^{NL}]$.
\begin{align}
L^{NL}_{c0}=& L_{c0}\left(\frac{a_m}{a_r}\right)^{2/3}, \nonumber \\ 
B^{NL}_0[L_{c0}^{NL}]=& B_0[L_{c0}]\left(\frac{a_m}{a_r}\right)^{-1/3}.
\label{BL_nlinear}
\end{align} 
Here $a_m$ represents the scale factor at the matter radiation equility. As discussed in \cite{sharma2017} there is no significant change in the comoving coherence length and field strength in the matter dominated era after $a_m$. Thus Eq.($\ref{BL_nlinear}$) gives reasonable estimates of the present day comoving field strength and corerealtion length. To estimate the maximum possible value of the magnetic field at different reheating scale, firstly we take the lowest possible scale of reheating ($5$MeV) allowed by the Big Bang Nucleosynthesis bound \cite{bbn}. We have also considered reheating scales around QCD phase transition, Electro-Weak phase transition and at $1000$ GeV. For each of these reheating temperatures we calculate the bound on the scale of inflation using Eq.($\ref{bound1}$). Further we calculate the magnetic field strength and its correlation length both assuming frozen field evolution Eq.($\ref{BL_linear}$) and with turbulent decay using Eq.($\ref{BL_nlinear}$). The results are given in Table $\ref{table1}$.

The bound obtained in Eq.($\ref{bound}$) suggests that as we increase the reheating scale ($T_r$), inflationary scale ($H_f$) decreases. Since reheating occurs after end of inflation, the above behaviour suggests that the highest possible reheating scale is $\approx 4000$ GeV for $\aa=2$. We consider several reheating scales below this highest possible reheating scale. If we consider reheating at $1000$ GeV, the maximum possible magnetic field strength is $7.1 \times 10^{-12} $ G and its correlation length is $0.03$ Mpc. We also calculate magnetic field strength and its correlation length for $T_r=100$ GeV, $T_r=150$ MeV and $T_r=5$ MeV and the results are shown in Table $\ref{table1}$. It is evident from the table that as we decrease the reheating temperature, the maximum possible magnetic field strength as well as its correlation length increases. Specifically, we have $B_0^{NL} \sim 3.9\times 10^{-11}$ G, 
$\sim 9.9 \times 10^{-10}$ G, $6.4 \times 10^{-9}$ G for respectively 
$T_r=100$ GeV, $T_r=150$ MeV and $T_r=5$ MeV 
and $L_{c0}^{NL} = 0.07$ Mpc, $0.6$ Mpc and $1.6$ Mpc respectively
for the same reheating temperatures.

In the above estimates, we have assumed that the EM energy density reaches a value equal to the energy density in the inflaton field at reheating. Suppose $(\rho_E+\rho_B)|_r = \epsilon \rho_{\phi}|_r$, then for a particular $T_r$, the above estimated magnetic field strength will be decreased by a factor of $\sqrt{\epsilon}$ but the correlation scale will remain the same. In this case, the maximum allowed value of reheating temperature is also decreased by a modest amount. For example with $\epsilon= 10^{-4}$, the maximum $T_r$ becomes $\approx 1300 $ GeV.

\section{$\gamma$-ray constraints}\label{gamma}
We now consider the constraints from the $\gamma$-ray observation. Non detection of GeV photons in Blazars observation by the Fermi telescope puts a lower bound on the magnetic field strength \cite{neronov, taylor2011}. Emitted photons of TeV energy from blazars interacts with the Extragalactic background light and gives rise to pair production. These produced particles interacts with the Cosmic Microwave Background and generate the GeV energy photons by inverse compton scattering. If a magnetic field is present in the intergalactic medium, it can affect the trajectories of these charge particle and observed surface brightness of GeV emission due to inverse compton scattering can be suppressed. Further, If we consider that size of secondary emitting region is larger than the point spread function of the telescope then this gives a lower bound of $B \sim 10^{-15}$ G strength at the coherence scale $\ge 1$ Mpc. If the flux suppression mechanism is due to the time delay of secondary emission then one gets a lower bound of $B \sim 10^{-17}$ G at the same coherence scale. Below 1 Mpc as the coherence length decreases the lower bound on the magnetic field strength increases as $L_c^{-1/2}$ in both the cases. Here $L_c$ is the 
comoving correlation length of the magnetic field.

\begin{figure*}
 \epsfig{figure=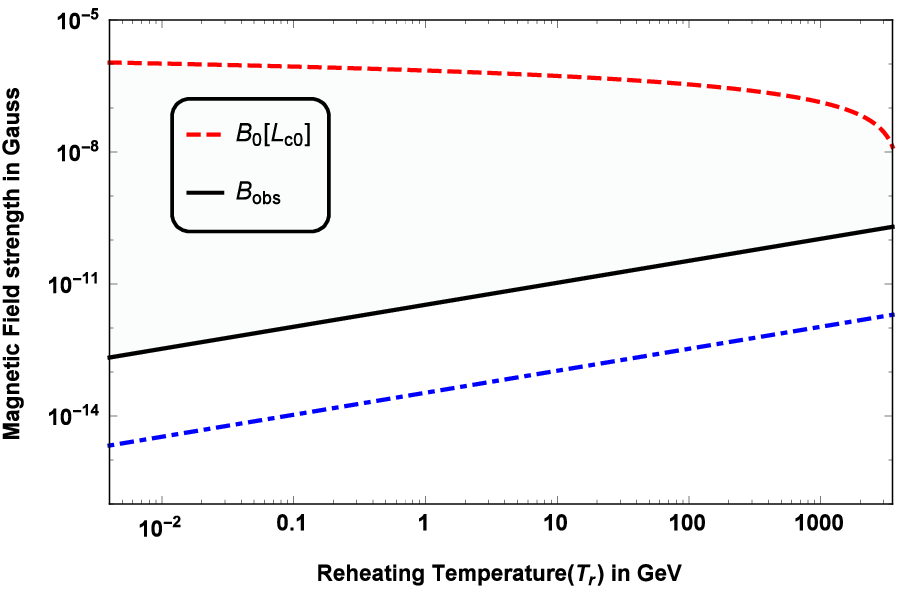,height=5cm,width=8cm,angle=0}
\epsfig{figure=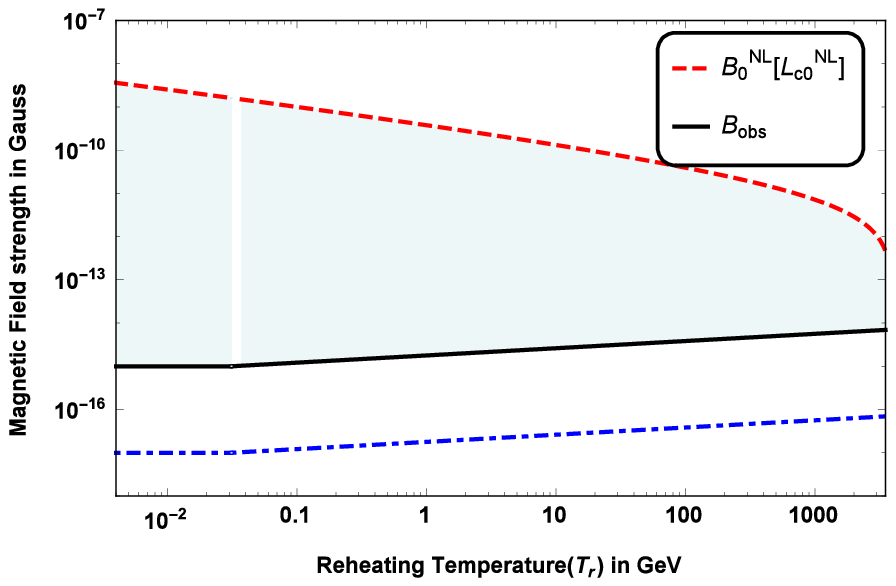,height=5cm,width=8cm,angle=0}
  \caption{{We have assumed ${\aa}=2$ in plotting these figures. The black curve and dotdashed blue curve in both the figures correspond to the lower bound on observed magnetic field strength constrained by the gamma ray observations for two different mechanism. These bounds are estimated at the correlation length of the generated magnetic field. The red dashed curve in the left panel represents the maximum magnetic field strength ($B_0[L_{c0}]$) that can be generated in our model by taking flux freezing evolution. While the red dashed curve in the right panel represents the maximum magnetic field strength ($B_0^{NL}[L^{NL}_{c0}]$) that can be generated by taking the nonlinear evolution of helical magnetic fields. The shaded region in both the figures represents all the allowed magnetic field strengths from $\gamma$-ray constraints.}}
\label{fig3}
\end{figure*}

In Fig.(\ref{fig3}), we have plotted the maximum possible magnetic field strength generated in our model for different reheating scales as a dashed red curve and the two bounds on the magnetic field strength obtained from the non-detection of $\gamma$-rays from blazars as respectively a solid black curve and as a dotdashed blue curve for the two different mechanism, respectively. In the left panel we have only considered the flux frozen evolution of the magnetic field after generation and estimated magnetic field strength and its correlation length. In right panel we have estimated the magnetic field strength and its correlation length by incorporating the non-linear evolution. It is evident from the figure that the generated magnetic field strength satisfies the $\gamma$-ray observation for all possible reheating scales in our model. The shaded region in the figure represents  the allowed values of magnetic fields from $\gamma$-ray observation.

In Fig.(\ref{fig3}), we have shown the constraints only for a certain range of the reheating temperature. BBN gives us a lower bound of 5 MeV for the scale of reheating. The reason for the upper bound is discussed in the previous
section. For $\aa=2$ case, it is at $4000$ GeV.
We see from Fig.(\ref{fig3}) that all our allowed models with
$T_r < 4000$ GeV, lead to magnetic fields well above the lower bound
required by the $\gamma$-ray observations.

\begin{widetext}
 \begin{center}
\begin{table}[h!]
\caption{Present day magnetic field strength and correlation length for different reheating scales ($T_r$)}
\begin{tabular}{|p{2 cm}|p{2 cm}|l|p{2 cm}|p{2 cm}|p{2 cm}|p{2 cm}|}
\hline
Scale of inflation (in GeV)
&Reheating Temperature $T_r$&${\aa}$&correlation length~$L_{c0}$ (in Mpc)&Magnetic field strength $B_0[L_{c0}]$(in G)&correlation length~$L^{NL}_{c0}$ (in Mpc)&Magnetic field strength $B_0^{NL}[L^{NL}_{c0}]$(in G)\\
\hline
$1.14\times10^{10}$
&$5$~~~~~MeV&2&$2.59\times10^{-5}$&$1.60\times10^{-7}$&1.62&$6.41\times10^{-9}$\\
\hline
$2.84\times10^{8}$&
$150$~~MeV&2&$6.46\times10^{-7}$&$9.34\times10^{-7}$&$0.58$&$9.90\times10^{-10}$\\
\hline
$3.88\times10^{5}$
&$100$~~GeV&2&$8.84\times10^{-10}$&$3.43\times10^{-7}$&$0.068$&$3.92\times10^{-11}$\\
\hline
$3.58\times10^{4}$
&$1000$~GeV&$2$&$8.84\times10^{-11}$&$1.35\times10^{-7}$&$0.032$&$7.12\times10^{-12}$\\
\hline
\end{tabular}
\label{table1}
\end{table}
\end{center}
\end{widetext}
\section{Conclusion}\label{conclusion}
We have studied here the generation of helical magnetic field during inflation. Generation of magnetic field within standard physics during inflation is not possible because of the conformal invariance of EM field. To generate magnetic field during inflation, we have adopted the Ratra model in which a coupling between EM field and inflaton field is assumed. We have added a parity violating term with the same coupling in our action to generate a helical magnetic field. However this model has the well known strong coupling and back-reaction problems during inflation. We have described these problems and attempted to resolve them by adopting a particular behaviour of coupling function $f$. In our model $f$ starts with a value of unity and increases during inflation so that there is no strong coupling and back-reaction problems. After inflation and before reheating it decreases such that it attains the pre-inflationary value at reheating to match with the observed coupling constant between charged fields and EM field. By demanding that there is no back-reaction of the generated fields post inflation, we get a bound on inflationary and reheating scales. For this type of evolution, the magnetic field spectrum at reheating is blue and can not be shallower than $d \rho _b/d \ln k \propto k^4$ spectra. This spectra is
obtained when $f\propto a^2$ during inflation. We have discussed this case in detail and estimate the magnetic field energy density and its correlation length for different reheating scales. If reheating happens at $100$ GeV then the comoving magnetic field strength is $3.4\times 10^{-7}$ G and its correlation length is $8.8\times 10^{-10}$ Mpc if we only consider the flux frozen evolution. Magnetic field strength and correlation length change to $3.9 \times 10^{-11}$ G and $0.07$ Mpc if we incorporate the non-linear evolution whereby the helical field decays due to the generated magnetohydrodynamics turbulence, conserving magnetic helicity. The generated magnetic field is almost fully helical.

The generated magnetic field strength at the end of inflation is larger compared to non-helical case considered in Ref.\cite{sharma2017}. Moreover the maximum possible reheating scale in the helical case is $\approx 4\times10^3$ GeV which was $\approx 10^4$ GeV in the non-helcial case. We have also shown that the generated magnetic field in our model satisfies the $\gamma$-ray constraints for all the allowed reheating scales.

The behaviour of coupling function which we have adopted, could be obtained in hybrid inflationary scenarios \cite{linde}. In hybrid inflation, one has two scalar fields with one dominating during inflation and providing the necessary condition for inflation and other field ends the inflation. We can consider our coupling function as a function of both the fields such that when first field rolls down the potential during inflation, $f$ increases and $f$ decreases when other field evolves.

To summarize, we have suggested a viable scenario for
inflationary generation of helical magnetic fields, which does
not suffer from the back reaction or strong coupling problems.
In our model the generated field is almost fully helical. 
As we increase the reheating scale both the 
magentic field strength and its correlation length decrease. However 
they
satisfy the $\gamma$-ray constraints for all the allowed values of reheating scales. 
The generated magnetic field strength
and its correlation length are larger for the helical case compared to the non-helical case. For a reheating scale at $100$ GeV the magnetic field strength and its correlation length were $6.8 \times 10^{-13} $ G and $ 7.3 \times 10^{-4}$ Mpc for the non-helical case \cite{sharma2017} but $3.9 \times 10^{-11} $ G and $0.07 $ Mpc for the helical case. 
Cosmic Microwave Background and structure formation have
mainly focused on nearly scale invariant spectra \cite{sethi2004, dner, pranjal2011, pranjal2013, planck_magnetic, pandey2014, kunze2014, kunze2015, chluba2015, kandu2016}.
It would be of interest to revisit these effects for the blue spectra (see for example \cite{wagstaff2015}) predicted by our consistent models of inflationary magnetogenesis.

\section*{Acknowledgments}
RS and TRS acknowledge the facilities at IUCAA Resource Center, University of Delhi as well as the hospitality and resources provided by IUCAA, Pune where part of this work has been done. RS acknowledges CSIR, India for the financial support through grant 09/045(1343)/2014-EMR-I. TRS acknowledges SERB for the project grant EMR/2016/002286.

\bibliographystyle{apsrev4-1}
\bibliography{references}
\end{document}